\documentclass[letterpaper,twocolumn]{jpsj3}
\usepackage{txfonts}
\usepackage{here}
\usepackage{graphicx}
\usepackage{dcolumn}
\usepackage{bm}
\usepackage{color}
\usepackage{ulem}
\title{Inhomogeneous Superconducting State Probed by $^{125}$Te NMR on UTe$_2$}
\author{Genki Nakamine$^1$\thanks{nakamine.genki.88v@st.kyoto-u.ac.jp}, Katsuki Kinjo$^1$, Shunsaku Kitagawa$^1$\thanks{kitagawa.shunsaku.8u@kyoto-u.ac.jp}, Kenji Ishida$^1$\thanks{kishida@scphys.kyoto-u.ac.jp}, Yo Tokunaga$^2$, Hironori Sakai$^2$, Shinsaku Kambe$^2$, Ai Nakamura$^3$, Yusei Shimizu$^3$, Yoshiya Homma $^3$, Dexin Li$^3$, Fuminori Honda$^3$, and Dai Aoki$^{3,4}$}
\inst{$^1$Depertment of Phisics, Kyoto University, Kyoto, 606-8502, Japan \\
$^2$ASRC, Japan Atomic Energy Agency, Tokai-mura, Ibaraki 319-1195, Japan \\
$^3$IMR, Tohoku University, Oarai, Ibaraki, 311-1313, Japan \\
$^4$University Grenoble, CEA, IRIG-PHERIQS, F-38000 Grenoble, France} 

\abst{UTe$_2$ is a recently discovered promising candidate for a spin-triplet superconductor.
In contrast to conventional spin-singlet superconductivity, spin-triplet superconductivity possesses spin and angular momentum degrees of freedom.
To detect these degrees of freedom and obtain the solid evidence of spin-triplet superconductivity in UTe$_2$, we performed $^{125}$Te-NMR measurement, which is sensitive to the local spin susceptibility at a nuclear site.
We previously reported that the shoulder signal appears in NMR spectra below the superconducting (SC) transition temperature $T_{\rm c}$ in $H \parallel b$, and that a slight decrease in the Knight shift along the $b$ and $c$ axes ($K_b$ and $K_c$, respectively) below $T_{\rm c}$ at a low magnetic field $H$.
Although the decrease in $K_c$ vanished above 5.5~T, the decrease in $K_b$ was independent of $H$ up to 6.5~T.   
To clarify the origin of the shoulder signal and the trace of the decrease in $K_b$, we compared the $^{125}$Te-NMR spectra obtained when $H~\parallel~b$ and $H~\parallel~c$ and measured the $^{125}$Te-NMR spectra for $H~\parallel~b$ up to 14.5~T.
The intensity of the shoulder signal observed for $H~\parallel~b$ has a maximum at $\sim 6$~T and vanishes above 10~T, although the superconductivity is confirmed by the $\chi_{\rm AC}$ measurements, which can survive up to 14.5~T (maximum $H$ in the present measurement).
Moreover, the decrease in $K_b$ in the SC state starts to be small around 7~T and almost zero at 12.5~T. This indicates that the SC spin state gradually changes with the application of $H$. 
Meanwhile, in $H~\parallel~c$, unexpected broadening without the shoulder signals was observed below $T_{\rm c}$ at 1~T, and this broadening was quickly suppressed with increasing $H$. 
We construct the $H$--$T$ phase diagram for $H~\parallel~b$ and $H~\parallel~c$ based on the NMR measurements and discuss possible SC states with the theoretical consideration.   
We suggest that the inhomogeneous SC state characterized by the broadening of the NMR spectrum originates from the spin degrees of freedom. }

\begin{document}
\maketitle
\section {\label{sec:Intro}Introduction}
Superconductors exhibit two types of spin states in general: spin-singlet ($S$ = 0) and spin-triplet ($S$ = 1).
For most superconductors, an $s$-wave pairing with a spin-singlet state and zero momentum results in a superconducting (SC) state with no degrees of freedom.
In contrast, the spin degrees of freedom remain in the spin-triplet state, and they possibly cause unusual behaviors. 
This includes the magnetic field ($H$)-boosted superconductivity, multiple SC phases, and rotation of the $\bm{d}$ vector, which is an SC order parameter for the spin-triplet pairing\cite{AokiJPSJRev2019, Mineev2017}.
However, there are few materials in which the spin-triplet state is considered to be realized.
In addition, the experimental observations for these phenomena are quite limited.
Some of the most promising candidates for spin-triplet superconductors are uranium (U)-based ferromagnetic (FM) superconductors UGe$_2$, URhGe, and UCoGe\cite{SaxenaNature2000, AokiNature2001, HuyPRL2007}, in which the ferromagnetism coexists with the superconductivity.
For these superconductors, many experimental and theoretical studies have suggested that spin-triplet superconductivity is induced by the Ising-type FM spin fluctuations\cite{AokiJPSJRev2019}.
One of the most convincing results is the $H$-boosted superconductivity caused by the development of the FM spin fluctuations, which has been observed in the vicinity of the $H$-induced quantum critical point for the U-based FM superconductors URhGe\cite{LevyScience2005} and UCoGe\cite{HuyPRL2008}. 
Therefore, research on the spin-triplet pairing state has been conducted for these superconductors.

Recently, the superconductivity of UTe$_2$ was discovered\cite{RanScience2019}; since then, UTe$_2$ has attracted significant attention because of its unique SC properties.
UTe$_2$ has the orthorhombic crystal structure with the space group $Immm$ (\#71, $D_{2h}^{25}$).
Its SC upper critical fields ($H_{\rm c2}$) for all three crystalline axes greatly exceed the Pauli limiting field\cite{RanScience2019, AokiJPSJ2019}; particularly for $H \parallel b$, the SC transition temperature $T_{\rm c}$ has an anomalous upturn at $\mu_0 H~=~15$~T, which is similar to the $H$-boosted behavior\cite{KnebeJPSJ2019}.
This $H_{\rm c2}$ behavior cannot be understood by the scenario in the spin-singlet superconductors, and it is reminiscent of the FM superconductors.
In addition to $H_{\rm c2}$, the Ising-type anisotropy of the static magnetic susceptibility\cite{RanScience2019, AokiJPSJ2019}, and the presence of the moderate Ising-type fluctuations\cite{TokunagaJPSJ2019} are similar to those observed in FM superconductors\cite{HardyPRL2005, HuyPRL2007, AokiJPSJ2009UCoGe, IharaPRL2010, HattoriPRL2012, TokunagaPRL2015, TokunagaPRB2016}. However, UTe$_2$ does not show any magnetic order at the ambient pressure\cite{RanScience2019}.
Therefore, it is suggested that UTe$_2$ is located in the vicinity of the FM quantum critical point and that the spin-triplet superconductivity mediated by the Ising-type FM spin fluctuations is realized in UTe$_2$.
Furthermore, multiple SC phases have been reported in the pressure ($P$)--temperature ($T$) phase diagram for UTe$_2$\cite{BraithwaiteComnPhys2019, RanPRB2020, ThomasSciAd2020}. Another SC phase is induced by applying pressure, and the evolution to more complex multiple SC phases in $H~\parallel~a$ under pressure was reported\cite{AokiJPSJ2020}.
These experimental results strongly suggest that UTe$_2$ is a spin-triplet superconductor with degenerate multiple SC order parameters.


To investigate the SC properties of UTe$_2$ from a microscopic point of view, we previously performed $^{125}$ Te-NMR measurements at the ambient pressure\cite{NakamineJPSJ2019, NakaminePRB2021}.
This is because the NMR measurement is sensitive to the changes in the local spin susceptibility at a nuclear site even in the SC state. 
When the $\bm{d}$ vector is pinned to one of the crystalline axis, Knight shift proportional to the spin susceptibility decreases in the SC state in $\bm{H} \parallel \bm{d}$, but remains unchanged in $\bm{H} \perp \bm{d}$.
Thus, the magnetic properties of the spin-triplet SC order parameters can be investigated. 
During our previous measurements\cite{NakaminePRB2021}, we focused on the reduction in the spin susceptibility in the SC state and measured the NMR Knight shift along the $b$ and $c$ axes ($K_b$ and $K_c$, respectively). 
$K_b$ and $K_c$ decrease slightly at 1~T, which indicates that the $\bm{d}$ vector has the $b$ and $c$ components.
This result limits the SC pairing symmetry in UTe$_2$ to $A_{u}$ or $B_{3u}$ in odd-parity states within the irreducible representations of $D_{2h}$.
We also reported the anisotropic response of the Knight shift reduction against $H$. 
The reduction in $K_b$ is nearly constant of $H$ up to 6.5~T; however, the reduction in $K_c$ decreases with increasing $H$. 
In addition, $K_c$ does not change even below $T_{\rm c}$ at 5.5~T.
This anisotropic response of the spin susceptibility is considered to be related to the SC pairing symmetry.

Although we have performed intensive research thus far, the following issues remain unresolved.
First, the origin of additional NMR signals in the SC state for $H~\parallel~b$ is unclear.
As previously reported\cite{NakaminePRB2021}, in addition to the shift of the main peak, additional shoulder signals appear in the SC-state NMR spectrum in $H~\parallel~b$.
Although the presence of additional signals suggests the appearance of another ordered state inside the SC phase, we cannot clarify the origin and characteristics from the previous experiments.
Second, it must be clarified whether $\bm{d}$-vector rotation occurs with a high $H$ along the $b$ axis.
The reduction in $K_b$ in the SC state implies that the SC state is unstable with a high $H$, which is due to the competition between the SC-condensation and the Zeeman energies.
In such a case, $\bm{d}$-vector rotation that is related to a change in the SC pairing symmetry is expected at a critical magnetic field $H_{\rm pin}$.
We estimated the $\mu_0H_{\rm pin} \sim$ 13~T from the decrease in $K_b$, as reported in our previous paper\cite{NakaminePRB2021}.

To solve these issues, in the present study, we performed precise NMR spectrum measurements up to 6.5 T in $H~\parallel~b$ and $H~\parallel~c$, and a high-$H$ NMR measurement up to 14.5 T in $H~\parallel~b$.
We found that the additional NMR signals appear well below $T_{\rm c}$ and the area fraction for those at 0.1~K against the main peak reaches a maximum at $\sim$ 6~T, and it becomes zero above 10~T.
Meanwhile, an unexpected large broadening of the NMR spectrum was observed in the SC state for $H~\parallel~c$, and it is quickly suppressed by $H$.
Furthermore, the reduction in $K_b$ starts to decrease at $\sim$ 7~T and it becomes zero at $\sim$ 12.5~T. This indicates that the $b$ components of the $\bm{d}$ vector disappear above 12.5~T.
Considering all these results, we discuss the possible SC states based on the phase diagram.

\section {\label{sec:Experimentals}Experimentals}

The single-crystal UTe$_2$, which is the same sample used in our previous study\cite{NakaminePRB2021}, was grown using the chemical transport method with iodine as a transport agent\cite{AokiJPSJproc2019}. 
Natural uranium and 99.9\% $^{125}$Te-enriched metals were used as the starting materials for the present sample. 
There are two crystallographically inequivalent Te sites, 4$j$ and 4$h$, with point symmetries $mm$2 and $m$2$m$, respectively.
We denote these sites as Te(1) and Te(2) according to the previous reports\cite{NakaminePRB2021}.
The measurements were performed using the single crystal with a size of $3.5 \times 0.7 \times 1.4$ mm$^3$. 
$T_{\rm c}$ was determined by the AC susceptibility that was measured by recording the resonance frequency of the NMR-tank circuit during cooling.
The frequency-swept $^{125}$Te~($I$~=~1/2, gyromagnetic ratio $^{125}\gamma_n/2\pi$~=~13.454~MHz/T)-NMR spectrum was obtained by the integration of the Fourier transform (FT) for a spin-echo signal that was observed after a radio frequency (RF) pulse sequence with a 3~kHz step for a fixed $H$.
The typical RF pulse length is 12~$\mu$s.
The NMR spectra are plotted against $K~\equiv~(f-f_0)/f_0$.
Here, $f$ is the NMR frequency, and $f_0$ is the reference frequency determined as $f_0~=~(\gamma_n/2 \pi)\mu_0H$.
The magnetic field was calibrated using $^{63}$Cu ($^{63}\gamma_n/2\pi$ =11.285 MHz/T) and $^{65}$Cu ($^{65}\gamma_n/2\pi$ =12.089 MHz/T) NMR signals from the NMR coil. 
In the present measurement, the errors of the NMR intensity and the linewidth are determined from the background noise amplitude and the resolution of FT signals, respectively.
All the NMR spectra in the SC state were recorded with the field-cooling process.
Low-temperature NMR measurements down to 80~mK were performed using a $^3$He-$^4$He dilution refrigerator, and the single-crystalline sample was immersed into the mixture to reduce the heating effect. 
In the low-$T$ NMR measurements, we reduced the power of the RF-pulses for the observation of the spin-echo signal as small as possible. 
Due to this effort, we detected the broadening of the NMR spectra below $T_{\rm c}$ originating from SC diamagnetism in all measurements.
We performed the measurements using two setups.
In the first setup (Setup~A), we used the SC split magnet, which generates a horizontal field up to 6.5 T with a single-axis rotator to apply $H$ exactly parallel to the $b$ axis or $c$ axis with the $a$ axis as the rotation axis.
This is the same setup as in the previous measurements\cite{NakaminePRB2021}.
The sample alignment and angle dependence of the NMR spectrum are reported in the supplemental materials in the previous report\cite{NakaminePRB2021}.   
In the second setup (Setup~B), we used the SC solenoid magnet, which generates a vertical field up to 14.5~T to apply a higher $H$ along the $b$ axis. 
In Setup~B, because the sample cannot be rotated in $H$, a misalignment with $H$ occurred, which is evaluated to be $3 \sim 13$ degrees tilted from the $b$ axis. 
The evaluation of the misalignment is discussed in a later section.

\section {\label{sec:results}Experimental results}
\subsection{Setup A: measurements with a rotator }
\begin{figure}[tb]
\begin{center}
\includegraphics[width=85mm]{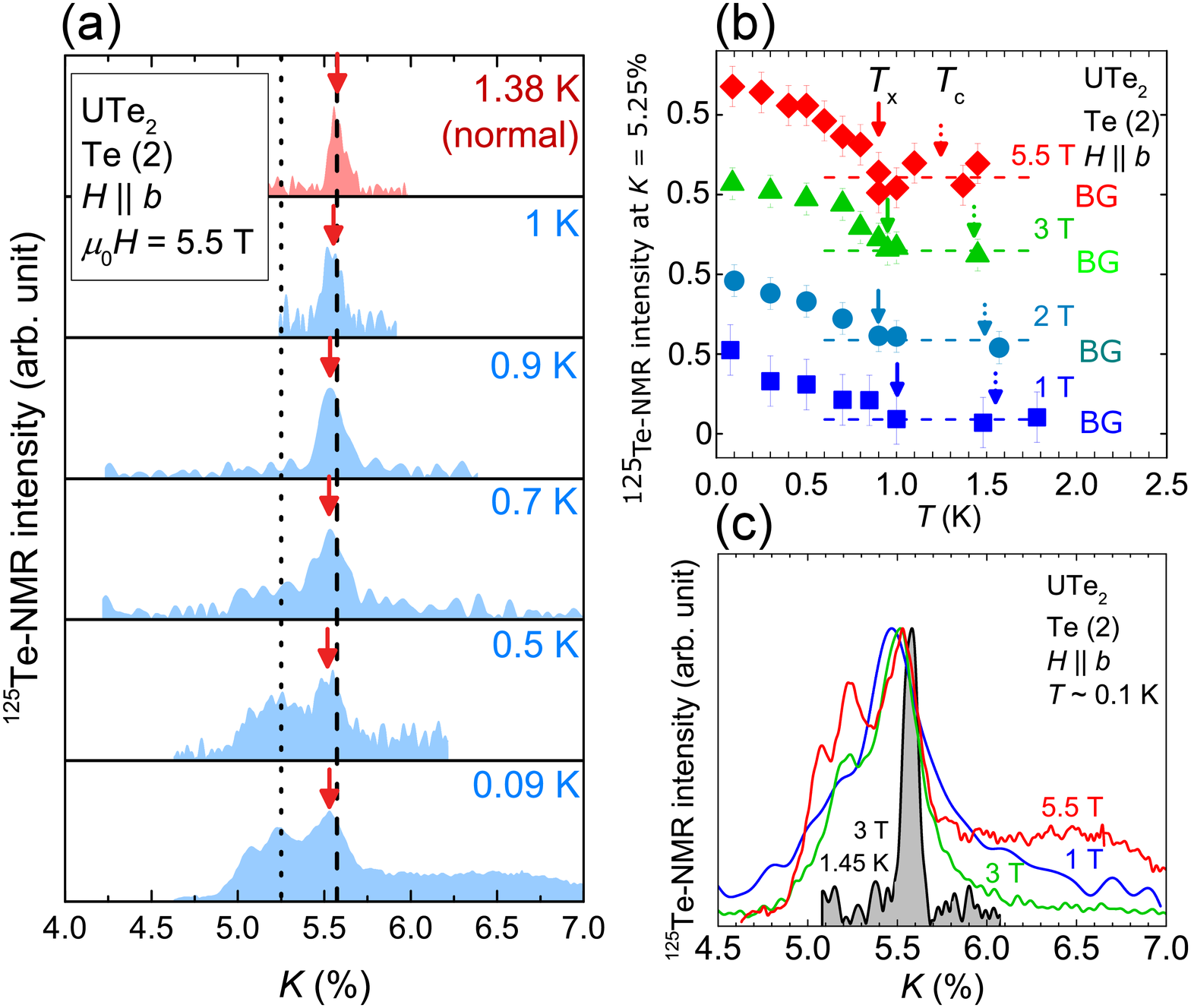}
\caption{\label{fig:fig1} (Color online) (a) $T$ variation of $^{125}$Te-NMR spectrum against $K~\equiv~(f-f_0)/f_0$ measured at 5.5~T along the $b$ axis.
The arrows indicate the peak of the ``main peak'' signal.
The dashed and dotted lines denote the Knight shift in the normal state and $K~=~5.25$\%, respectively.
(b) Temperature dependence of the NMR intensity at $K = 5.25$\% normalized by the main-peak intensity at $\mu_0H = 1, 2, 3,$ and 5.5~T.
Background (BG) noise level is shown by dotted lines. 
To avoid overlapping data, the offset value (0.5) is added.
The SC transition temperatures $T_{\rm c}$ obtained by the AC susceptibility measurement are shown by the dotted arrows.
The solid arrows indicate the characteristic temperature $T_x$ at which the intensity starts to increase upon cooling.
(c) $^{125}$Te-NMR spectrum at the lowest temperature at $\mu_0~H~=~1, 3,$ and 5.5~T along the $b$ axis.
For comparison purposes, the normal-state NMR spectrum at 3~T is also shown.
}
\end{center}
\end{figure}

First, we summarized the experimental results in Setup~A.
Figure~\ref{fig:fig1}(a) shows the $^{125}$Te-NMR spectrum at several temperatures measured at 5.5~T in $H~\parallel~b$.
The peak shown by the arrows shifts to a lower $K$ direction below $T_{\rm c}$. 
This indicates a decrease in the spin susceptibility by forming Cooper pairs.
In addition, we found that additional signals appear at low temperatures.
Hereafter, we refer to the peak of the NMR spectrum and the additional signal that is observed at low temperatures as the ``main peak'' and ``shoulder signal", respectively. 
To investigate the origin of the shoulder signal appearing in the SC state upon cooling, the temperature dependence of the NMR intensity at $K~=~$5.25\% [shown by the dotted line in Fig.~\ref{fig:fig1}(a)], which is normalized by the main peak intensity, is plotted in Fig.~\ref{fig:fig1}(b).
We can define the characteristic temperature $T_x$, at which the intensity at $K~=~$5.25\% starts to increase. 
$T_x$ shown by the solid arrows in Fig.~\ref{fig:fig1}(b) is $\sim 1$~K and is almost independent of $H$ in the measured $H$ region, which is different from $T_{\rm c}$.
These results indicate the presence of another phase below $T_x$.
The difference between $T_{\rm c}$ and $T_x$ is $\sim$~0.6~K at 1~T, which is significantly larger than the separation in the double transition that is observed by a specific heat at zero fields ($\sim$~0.2~K)\cite{ThomasSciAd2020, hayesArXiv2020}.
It is noted that the shoulder signal has a characteristic structure.
There are two peaks at the lower $K$ side and a broad peak that tails to a higher $K$ than that of the main peak, as shown in Fig.~\ref{fig:fig1}(c).
This indicates that the spin susceptibility is inhomogeneously distributed in the low-$T$ phase.
The area fraction of the shoulder signal at 0.1~K against the main peak increases with increasing $H$ up to 5.5~T, although the main peak still remains.
Since we require higher-$H$ NMR measurements than 6.5 T to clarify how the area fraction of the shoulder signal changes against $H$, the measurement was performed using Setup~B as described in section~3.2.

\begin{figure}[tb]
\begin{center}
\includegraphics[width=85mm]{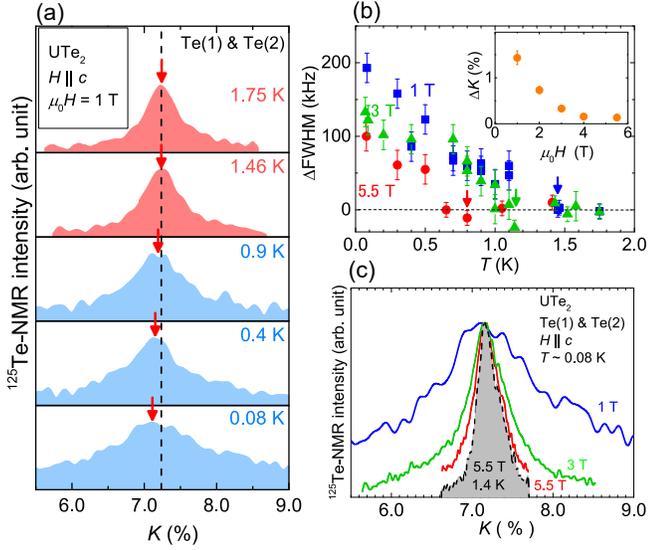}
\caption{\label{fig:fig2} (Color online) (a) $T$ variation of $^{125}$Te-NMR spectrum against $K$ measured at 1 T along the $c$ axis. 
The dashed line denotes the Knight shift in the normal state.
(b) Temperature dependence of the FWHM for the NMR spectrum from which the normal-state value is subtracted ($\Delta$FWHM) at $\mu_0H = 1, 3,$ and 5.5 T. 
The arrows show $T_{\rm c}$s. 
The inset shows the magnetic field dependence of the $\Delta$FWHM at the lowest temperature divided by the measured frequency ($\Delta K$).   
(c) $^{125}$Te-NMR spectrum at the lowest temperature at $\mu_0H =1, 3,$ and 5.5 T along the $c$ axis.
For comparison purposes, the normal-state NMR spectrum at 5.5~T is also shown.}
\end{center}
\end{figure}

We also performed $^{125}$Te-NMR measurements in $H~\parallel~c$ by Setup~A. 
Figure~\ref{fig:fig2}(a) shows the $^{125}$Te-NMR spectrum against $K$ at several temperatures measured at 1~T in $H~\parallel~c$.
Although a single symmetric peak was observed in the normal state, this peak consists of two NMR signals arising from the Te(1) and Te(2) sites because of the closeness of the Knight shift at the two sites when $H$ is parallel to the $c$ axis\cite{TokunagaJPSJ2019, NakaminePRB2021}.
The symmetric NMR spectrum becomes broader in the SC state without showing any shoulder peak structure observed in $H~\parallel~b$, and it slightly shifts to a lower $K$ direction, as we previously reported \cite{NakaminePRB2021}.
The temperature dependence of the full width at the half maximum (FWHM) of the NMR spectrum, from which the normal-state value is subtracted ($\Delta$FWHM), in the frequency units at 1, 3, and 5.5~T are shown in Fig.~\ref{fig:fig2}(b).
Although the $\Delta$FWHM for all the $H$ values increases below $T_{\rm c}$, the $\Delta$FWHM at the lowest $T$ becomes smaller with increasing $H$, as shown in the inset of Fig.~\ref{fig:fig2}(b).
It is noted that the broadening at the lowest $T$ is one order of magnitude larger than the broadening expected from the conventional SC diamagnetic effect.
The broadening of the NMR spectrum by the conventional SC diamagnetic effect is of the order of the SC lower critical field $\mu_0H_{\rm c1} \sim 1.2$~mT\cite{NakaminePRB2021}. 
This gives the upper limit of the $^{125}$Te-NMR spectrum broadening $\sim$16~kHz.
Therefore, it is concluded that the broadening of the $^{125}$Te-NMR in $H~\parallel~c$ is not simply due to the conventional SC diamagnetic effect. Instead, it originates from the unconventional SC properties in UTe$_2$.
The unexpected large broadening of the NMR spectrum in the SC state suggests that the spin susceptibility is inhomogeneously distributed also in $H \parallel c$.
It is also noteworthy that the magnetic field dependence of the broadening of the NMR spectrum is different between $H~\parallel~b$ and $H~\parallel~c$.
The NMR spectrum at the lowest temperature, which is shown against $K$, quickly becomes narrower with an increase in $H$ when $H~\parallel~c$, as shown in the inset of Fig.~\ref{fig:fig2}(b) and Fig.~\ref{fig:fig2}(c).
Although the magnetic field at which the broadening behavior disappears is ambiguous in $H \parallel c$, it is roughly estimated as $\sim$~3 T above which the $\Delta K$ is almost independent of $H$ as shown in the inset of Fig.~\ref{fig:fig2}(b).
Here, $\Delta K$ is the $\Delta$FWHM at the lowest temperature divided by the measured frequency. 
In contrast, the broadening behavior becomes more obvious with an increase in $H$ when $H~\parallel~b$, as shown in Fig.~\ref{fig:fig1}(c).
This indicates that the spin-susceptibility distributed signal, observed in the low fields, is suppressed in $H \parallel c$, but it seems to be enhanced in $H \parallel b$.

\subsection{Setup B: high-$H$ measurements in $H \parallel \sim b$ }
\begin{figure}[tb]
\begin{center}
\includegraphics[width=8cm,clip]{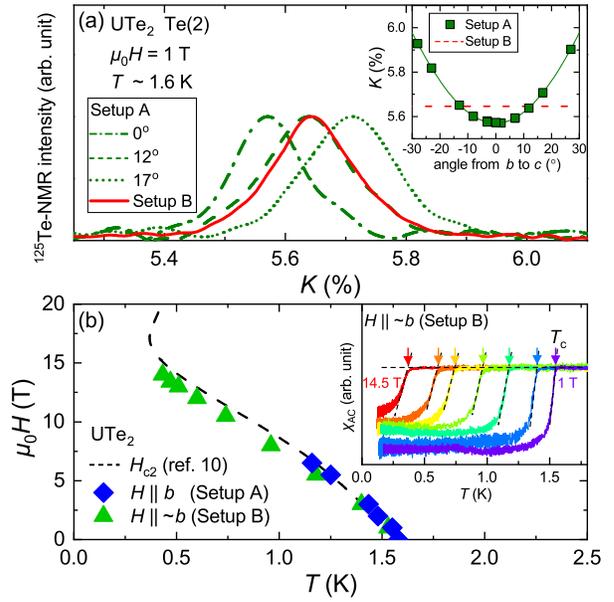}
\end{center}
\caption{\label{fig:fig3} (Color online) (a)$^{125}$Te-NMR spectrum in Setup~B together with those in Setup~A at several field angles from the $b$ to $c$ axis at 1~T and 1.65~K.
(Inset)Angular dependence of the Knight shift in Setup~A (squares) together with the Knight shift in Setup~B (dashed line). 
(b)The $H$--$T$ phase diagram for $H \parallel b$ in Setup~A and for $H \parallel \sim b$ in Setup~B.
The broken curves represent $H_{\rm c2}$ from the reference\cite{KnebeJPSJ2019}.
(Inset)Temperature dependence of $\chi_{\rm AC}$ at several $H$ in Setup~B.
$T_{\rm c}$s, which are represented by the arrows, were determined by the intersection point of the slopes of the SC diamagnetism and the normal state.
}

\end{figure}

To investigate the origin of the shoulder signal and the SC properties for a high $H$, we performed the NMR measurements using Setup~B.
As described in Section 2, there is a small sample misalignment in Setup~B because the field direction cannot be controlled.
Figure~\ref{fig:fig3}(a) shows the $^{125}$Te-NMR spectrum in Setup~B along with those in Setup~A at several field angles from the $b$ to $c$ axis at 1~T and 1.65~K.
The NMR spectrum of Setup~B almost overlaps with the NMR spectrum at 12$^\circ$ in Setup~A. 
From the comparison between the Knight shift in Setup~B and the angular dependence of the Knight shift in Setup~A, we estimated that the tilted angle is 12.5$^\circ$ from the $b$ axis to the $c$ axis when we assume that the sample misalignment originates from only the $b$--$c$ rotation as shown in the inset of Fig.\ref{fig:fig3}~(a).
We can also estimate that the tilted angle is $\sim$~3$^\circ$ from the $b$ axis to the $a$ axis because there is a possibility of sample misalignment with the $b$--$a$ rotation.
Here, we used the Knight shift along the $a$ axis $K_a$~$\sim$~40\% that was obtained by extrapolating the temperature dependence of $K_a$ from 20~K to 1.6~K.
It has been reported that the NMR signal for $H~\parallel~a$ disappears below 20~K due to the divergence of $1/T_2$\cite{TokunagaJPSJ2019}; thus, $K_a$ at low temperatures could not be measured.
To check the effect of the misalignment on the superconductivity, we measured the magnetic field dependence of $T_{\rm c}$ in Setup~B from the $T$-scan of $\chi_{\rm AC}$, as shown in the inset of Fig.~\ref{fig:fig3}(b).
$T_{\rm c}$ was determined by the intersection point of the slope of the SC diamagnetism and the normal state.
The obtained magnetic field dependence of $T_{\rm c}$ is plotted in Fig.~\ref{fig:fig3}(b) with that measured in Setup~A, where $H$ is exactly parallel to the $b$ axis. 
Although $T_{\rm c} (H)$ is slightly reduced from that in $H~\parallel~b$, the shape of the phase diagram is almost the same up to 14.5~T. This indicates that the effect of the misalignment is small for the superconductivity.

\begin{figure}[tb]
\begin{center}
\includegraphics[width=8.3cm,clip]{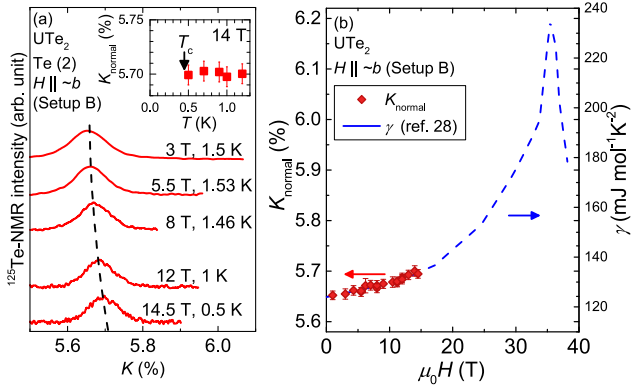}
\end{center}
\caption{\label{fig:fig4}  (Color online) (a)$^{125}$Te-NMR spectrum at several $H$ in the normal state. 
The dashed line is a guide for the eye. 
(Inset)Temperature dependence of $K_{\rm normal}$ at 14~T.
(b) Magnetic field dependence of $K_{\rm normal}$ and the Sommerfeld coefficient $\gamma$\cite{ImajoJPSJ2019}.
}
\end{figure}

Next, we show the magnetic field dependence of the magnetic properties in the normal state.
Figure~\ref{fig:fig4}(a) shows the $^{125}$Te-NMR spectrum at several $H$ values in the normal state.
A peak in the NMR spectrum gradually shifts to a higher $K$ direction with increasing $H$. 
The Knight shift in the normal state, $K_{\rm normal}$ is almost independent of $T$ below 2~K, as shown in the inset of Fig.~\ref{fig:fig4}(a). 
Because $K_{\rm normal}$ is related to the Pauli paramagnetism, the increase in $K_{\rm normal}$ against $H$ can be interpreted as an increase in the density of the states, which is consistent with the previous bulk measurements\cite{ImajoJPSJ2019, MiyakeJPSJ2019}.
The magnetic field dependence of $K_{\rm normal}$ and the Sommerfeld coefficient $\gamma$\cite{ImajoJPSJ2019} is presented in Fig.~\ref{fig:fig4}(b).
The observed increase in $K_{\rm normal}$ was scaled with $\gamma$.

\begin{figure}[tb]
\begin{center}
\includegraphics[width=70mm]{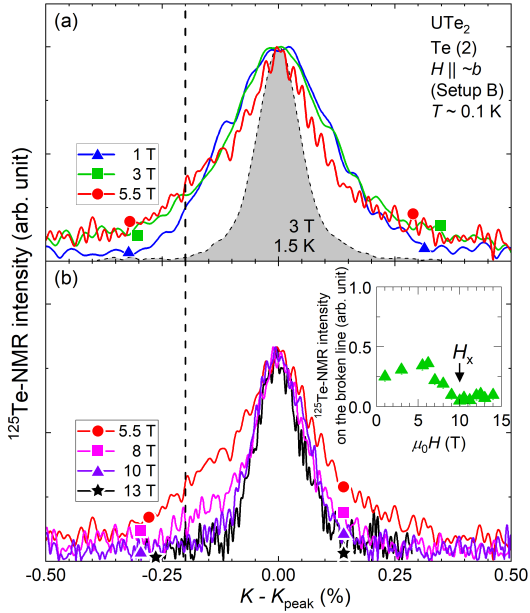}
\caption{\label{fig:fig5}  (Color online) $^{125}$Te-NMR spectrum at $\sim$0.1~K in (a) $\mu_0 H \leq$ 5.5~T and (b) $\mu_0 H \geq 5.5$~T in Setup~B.
The dashed line denotes the position of $K - K_{\rm peak}~=~-0.2$\%.
(Inset) Magnetic field dependence of the NMR intensity at $K - K_{\rm peak}~=~-0.2$\% measured at 0.1~K. 
}
\end{center}
\end{figure}

The shoulder signal in the SC state displays peculiar behavior.
Figure~\ref{fig:fig5}(a) shows the $^{125}$Te-NMR spectrum measured at $\sim 0.1$~K up to 5.5~T in Setup~B.
As previously mentioned and illustrated later, the Knight shift in the normal and SC states depend on $H$. Thus, to simply compare the NMR line shape, in which the NMR spectra are normalized with the maximum peak, it is plotted against $K-K_{\rm peak}$, where $K_{\rm peak}$ is the Knight shift at the maximum peak of the NMR spectrum.
Although the peaks were located at slightly different positions due to the misalignment from the $b$ axis, the shoulder signal observed in Setup~A was also observed in Setup~B.
Therefore, we could investigate the $H$ evolution of the shoulder signal in the high $H$ region.
Figure~\ref{fig:fig5}(b) shows the $^{125}$Te-NMR spectrum that was measured at $\sim 0.1$~K above 5.5~T. This is plotted in the same manner as Fig.~\ref{fig:fig5}(a).
As illustrated, the intensity of the shoulder signal decreases with increasing $H$, and it almost vanishes at approximately 10~T.
We plotted the magnetic field dependence of the intensity at $K-K_{\rm peak}~=~-0.2$\%, as shown by the dotted line in Fig.~\ref{fig:fig5}(b) at $\sim 0.1$~K in the inset of Fig.~\ref{fig:fig5}(b).
From the figure, we can determine the critical field $\mu_0 H_x$ where the shoulder signal disappears to be 10~T.

\begin{figure}[tb]
\begin{center}
\includegraphics[width=85mm]{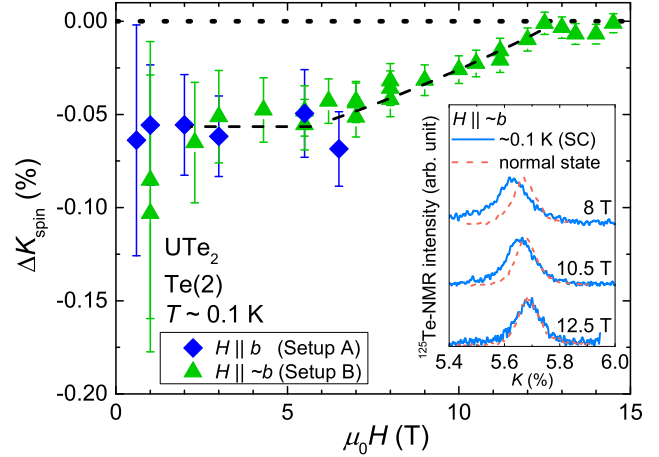}
\caption{\label{fig:fig6}
 (Color online) Magnetic field dependence of the change in the spin component of the Knight shift in the SC state, $\Delta K_{\rm spin}$ in Setup~A and B.
The dotted line indicates $\Delta K_{\rm spin}$ = 0.
The dashed lines are a guide for the eye.
(Inset) $^{125}$Te-NMR spectrum at $\sim 0.1$~K (SC state) measured at 8, 10.5, and 12.5~T in Setup~B.
For comparison purposes, the normal-state NMR spectrum is also shown.
}
\end{center}
\end{figure}

In addition to $H_{x}$, we also investigated the critical field at which the decrease in $K_b$ becomes zero.
In the inset of Fig.~\ref{fig:fig6}, the NMR spectrum at $\sim 0.1$~K (SC state) is compared with the normal-state NMR spectrum at 8, 10.5, and 12.5~T.
While the NMR spectrum at 8~T and 10.5~T shifts to a lower $K$ direction in the SC state, no spectrum shift was observed between the normal and SC states at 12.5~T.
For the quantitative discussion, we evaluated the change in the spin component of the Knight shift in the SC state, $\Delta K_{\rm spin}$, which is described as 
$$\Delta K_{\rm spin} = K(T \sim 0.1 \, {\rm K}) - K_{\rm normal} - \Delta K_{\rm dia},$$
and we plotted the magnetic field dependence of $\Delta K_{\rm spin}$ in Fig.~\ref{fig:fig6}.
Here, $K(T)$ is the Knight shift at $T$, and $\Delta K_{\rm dia}$ is the contribution from the conventional SC diamagnetism.
$\Delta K_{\rm dia}$ was estimated from the formula suggested by de Gennes\cite{deGennesSC}, which is expressed as 
$$\Delta K_{{\rm dia}} = - \frac{H_{\rm c1}}{H} \frac{\ln \left( \frac{\beta d}{\sqrt{e} \xi} \right) }{\ln{\kappa}}.$$
As shown in Fig.~\ref{fig:fig6}, the obtained $\Delta K_{\rm spin}$ is almost independent of $H$ up to 7~T, which is consistent with the previous results\cite{NakaminePRB2021}.
Meanwhile, the absolute value of $\Delta K_{\rm spin}$ decreases gradually above 7~T and becomes almost zero above $\sim$12.5~T.
We comment that the intensity of the shoulder signal starts to decrease at around 7~T, indicating that the SC state might be gradually changed from this field.
It is noted that the superconductivity survives at least up to 14.5~T because the SC transition was confirmed by the $\chi_{\rm AC}$ measurement and the broadening of the NMR spectrum.
The zero value of $\Delta K_{\rm spin}$ implies that the $\bm{d}$ vector has no $b$ component above 12.5~T.

\section{Discussion}
Here, we discuss how the $\bm{d}$ vector changes against $H$ along the $b$ axis.
In a previous report\cite{NakaminePRB2021}, we pointed out that the low-$H$ SC state is unstable; moreover, it should change in the high-$H$ region in $H~\parallel~b$ because $K_b$ decreases in the SC state, and the competition between the SC-condensation and Zeeman energies is expected.
We estimated the critical magnetic field $\mu_0H_{\rm pin}$ to be $\sim$ 13~T from the decrease in $K_b$, which is consistent with the observed critical field in which the decrease in $K_b$ becomes zero.
The observed magnetic field dependence of $\Delta K_{\rm spin}$ is gradual, not drastic.
One possibility is that the SC pairing symmetry below and above 12.5~T is the same representation in $H~\parallel b$. Therefore, the change in the SC state is a crossover rather than a first-order phase transition, which gives strong constraints for the SC pairing symmetry.

Next, we discuss the low-$T$ SC state, in  which the sharp signal arising from the high-$T$ SC phase and the broad spin-susceptibility-distributed signal (the shoulder signal in $H \parallel b$ and the unexpected broad signal in $H \parallel c$) are observed.
One of the characteristic features of the low-$T$ SC state is that the broad signal exists only below $T_c$ ($H$), indicating that this signal is closely related to the superconductivity.
The other is the anisotropic field response of the spectrum broadening.
The NMR spectrum at the lowest temperature quickly becomes narrower with an increase in $H$ when $H~\parallel~c$, although the broadening behavior becomes more obvious with an increase in $H$ and disappears at 10~T when $H~\parallel~b$.
These results exclude the possibility of the magnetic origin for the spectrum broadening. 
Therefore, one possible scenario to interpret the origin of the broad signal is that other SC states with different SC pairing symmetries from that in the high-$T$ SC phase are realized in the specific regions of the sample.
($i.e.$ the broad signal arises from the low-$T$ SC phase.)
Since the NMR spectrum is extremely broad, a spatially distributed order parameter or multi-component order parameter generating the internal field are expected for the low-$T$ SC phase. 
As only the high-$T$ SC phase exists between $T_c$ and $T_x$, the low-$T$ SC phase is expected to grow to replace a part of high-$T$ SC phase.
In general, such coexistence is induced by disorders or impurities.
However, the spectrum in the normal state is very sharp as shown in Figs.~\ref{fig:fig1}(a) and (c), and there are no indications of the presence of impurities or disorders.
Therefore, it is considered that the high-$T$ and low-$T$ SC phases are highly degenerated, and the ground state can change with a slight difference in the local situation.

\begin{figure}[tb]
\begin{center}
\hspace{5mm}
\includegraphics[width=80mm]{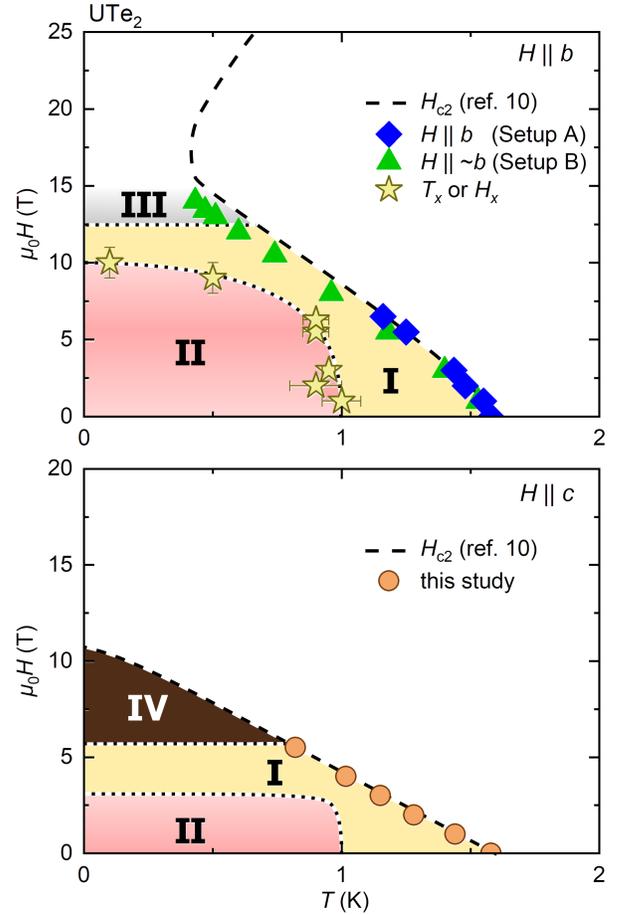}
\caption{\label{fig:fig7} (Color online) Summarized $H$--$T$ phase diagram for (a) $H~\parallel~b$ and (b) $H~\parallel~c$.
The diamonds, triangles and circles represent $H_{\rm c2}$ determined by $\chi_{\rm AC}$ in this study.
The broken curves represent $H_{\rm c2}$ from reference\cite{KnebeJPSJ2019}.
The stars represent $T_{x}$ and $H_{x}$ as described in the text.
The SC phase can be distinguished into four regions based on the NMR results.
Region boundaries are indicated by dotted lines.
Each region in the phase diagrams is numbered as I$\sim$\mbox{I\hspace{-.1em}V}. 
}
\end{center}
\end{figure}

We now summarize the $H$--$T$ phase diagram for $H~\parallel~b$ and $H~\parallel~c$ in UTe$_2$ in Figs.~\ref{fig:fig7} (a) and (b), respectively.
Based on the present results, we can distinguish the SC phase into four regions, and number each region in the phase diagrams as I$\sim$ \mbox{I\hspace{-.1em}V}.
In region I, only the main peak was observed, and $K_b$ and $K_c$ decrease in the SC state. 
This corresponds to the high-$T$ SC phase and indicates the single-phase SC state with the nonzero $b$ and $c$ components in the $\bm{d}$ vector ($\bm{d_b}$ and $\bm{d_c}$, respectively).
Considering the $D_{2h}$ point group symmetry, the $A_{u}$ or $B_{3u}$ state is a candidate for SC pairing symmetry, which is consistent with the theoretical suggestion\cite{IshizukaPRL2019, IshizukaPRB2021, ShishidouPRB2021}.
In region \mbox{I\hspace{-.1em}I}, the shoulder signal appears in addition to the main peak in $H~\parallel~b$.
The region \mbox{I\hspace{-.1em}I} corresponds to the low-$T$ SC state where the high-$T$ SC phase and the low-$T$ SC phase coexist as discussed above.
The determination of the boundary of the region \mbox{I\hspace{-.1em}I} for $H \parallel c$ is ambiguous.
We consider that region \mbox{I\hspace{-.1em}I} in $H \parallel c$ appears also below 1 K from the constraint that the transition between region I and region \mbox{I\hspace{-.1em}I} occurs at the same temperature at the zero field. 
In region \mbox{I\hspace{-.1em}I\hspace{-.1em}I}, $K_b$ does not decrease in the SC state. 
This indicates that the SC state with $\bm{d} \perp \bm{b}$ is realized.
Meanwhile, $K_c$ does not decrease in the SC state in region \mbox{I\hspace{-.1em}V}, indicating that the SC state with $\bm{d} \perp \bm{c}$.
Because the change in $\Delta K_{\rm spin}$ against $H$ is gradual in $H \parallel b$ and $H \parallel c$, it seems that the transitions between region I and \mbox{I\hspace{-.1em}I\hspace{-.1em}I} and between region I and \mbox{I\hspace{-.1em}V} are crossovers.
Since no phase transitions were detected between the different regions with other experimental techniques, a further experimental confirmation is desired.

Finally, we discuss a possible SC pairing symmetry in each region that is based on the present experimental results and the theoretical considerations.
The Table~\ref{tab:tab1} represents the character tables for point groups (a) $D_{2h}$, (b) $C^b_{2h}$, and (c) $C^c_{2h}$\cite{IshizukaPRL2019}.
Here, the $D_{2h}$ symmetry is reduced to $C_{2h}^{b}$ in $H~\parallel~b$ or $C_{2h}^{c}$ in $H~\parallel~c$.
The SC pairing symmetry is classified into the one of the irreducible representations listed on the tables, and the $\bm{d}$ vector can be represented by the corresponding basis function.
\begin{table}[htb]
\begin{center}
\caption{\label{tab:tab1}Character tables for point groups (a) $D_{2h}$, (b) $C^b_{2h}$, and (c) $C^c_{2h}$\cite{IshizukaPRL2019}. Only odd-parity irreducible representations (IR) are listed. 
The notation $x, y, z$ in the reference\cite{IshizukaPRL2019} correspond to $a, b, c$ in the table.}
\vspace{5mm}
\scalebox{0.85}[1]{ 
  \begin{tabular}{cccccccccc}\hline \hline 
 \multicolumn{10}{c}{(a) $D_{2h}$ (zero magnetic field)} \\ \hline 
IR & $E$ & $C_{2c}$ & $C_{2b}$  & $C_{2a}$ & $I$ & $\sigma_c$ & $\sigma_b$ & $\sigma_a$ & Basis functions \\ \hline
$A_u$ & 1 & 1 & 1 & 1 & -1 & -1 & -1 & -1 & $k_a \hat{a}$, $k_b \hat{b}$, $k_c \hat{c}$ \\
$B_{1u}$ & 1 & 1 & -1 & -1 & -1 & -1 & 1 & 1 & $k_b\hat{a}$, $k_a\hat{b}$ \\
$B_{2u}$ & 1 & -1 & 1 & -1 & -1 & 1 & -1 & 1 & $k_a \hat{c}$, $k_c \hat{a}$ \\
$B_{3u}$ & 1 & -1 & -1 & 1 & -1 & 1 & 1 & -1 & $k_c \hat{b}$, $k_b \hat{c}$ \\ \hline \hline
\\ \\
  \end{tabular}
 }
 \scalebox{0.95}[1]{ 
\begin{tabular}{ccccccc}\hline \hline
 \multicolumn{7}{c}{(b) $C_{2h}^b$ ($H \parallel b$)} \\ \hline 
 IR & (IR)$\uparrow D_{2h}$ & $E$ & $C_{2b}$ & $I$  & $\sigma_b$ & Basis functions \\ \hline
$A_u$ & $A_u + B_{2u}$& 1 & 1 & -1 & -1 & $k_a \hat{a}$, $k_b \hat{b}$, $k_c \hat{c}$, $k_a \hat{c}$, $k_c \hat{a}$   \\
$B_u$ & $B_{1u} + B_{3u}$ & 1 & -1 & -1 & 1  & $k_b\hat{a}$, $k_a\hat{b}$, $k_c \hat{b}$, $k_b \hat{c}$ \\ \hline \hline
\\ \\
 \end{tabular}
 }
 \\ 
 \scalebox{0.95}[1]{ 
 \begin{tabular}{ccccccc}\hline \hline
 \multicolumn{7}{c}{(c) $C_{2h}^c$ ($H \parallel c$)} \\ \hline 
 IR & (IR)$\uparrow D_{2h}$ & $E$ & $C_{2c}$ & $I$  & $\sigma_c$ & Basis functions \\ \hline
$A_u$ & $A_u + B_{1u}$& 1 & 1 & -1 & -1 & $k_a \hat{a}$, $k_b \hat{b}$, $k_c \hat{c}$, $k_b \hat{a}$, $k_a \hat{b}$   \\
$B_u$ & $B_{2u} + B_{3u}$ & 1 & -1 & -1 & 1  & $k_a \hat{c}$, $k_c \hat{a}$, $k_c \hat{b}$, $k_b \hat{c}$  \\ \hline \hline

 \end{tabular}
  }
  \end{center}
\end{table}
As mentioned above, the plausible SC pairing symmetry in region~I is $A_u$ or $B_{3u}$. 
In region III(IV), the $K_b$($K_c$) does not decrease and thus, the $B_{2u}(B_{1u})$ state without $d_b$($d_c$) component is expected.
From the crossover behavior of the boundary between region I and region III(IV), the SC pairing symmetry of region I should be the same irreducible representation as that of region III(IV) under $H \parallel b$ ($H \parallel c$).
As shown in Table~\ref{tab:tab1}, the $A_u$ state which is in the same representation as $B_{2u}$ state in $C_{2h}^b$ and as $B_{1u}$ state in $C_{2h}^c$ is more suitable than the $B_{3u}$ state for the SC pairing symmetry of region~I. 
To determine the SC pairing symmetry at region I, the Knight shift measurement in $H \parallel a$ is crucial.

A recent theoretical calculation that is based on a periodic Anderson model reveals that the $A_{u}$ and $B_{3u}$ states are almost degenerate and the dominant component of the $\bm{d}$-vector for the $A_{u}$($B_{3u}$) state is $\bm{d_b}$($\bm{d_c}$)\cite{IshizukaPRB2021}. 
This is consistent with our experimental results. 
According to the calculation, it seems that the degeneracy of the $A_{u}$ and $B_{3u}$ states results in the coexistence of the high-$T$ and low-$T$ SC phases expected in region \mbox{I\hspace{-.1em}I}.
In region \mbox{I\hspace{-.1em}I}, the broad signal is quickly suppressed against $H$ along the $c$-axis, indicating that the low-$T$ SC phase includes $B_{3u}$ component with a large $\bm{d_c}$.
We suggest that the $\alpha A_u + \beta B_{3u}$ state, whose complex coefficients $\alpha, \beta$ are changed depending on the local situations, can be consistent with the characteristic NMR spectrum observed in region \mbox{I\hspace{-.1em}I}.

For multi-component SC state, the time-reversal symmetry-breaking SC state originating from a chiral or a nonunitary state is expected.
Although our measurements were performed under fields, our results are possibly related to the recent results with the polar Kerr effect\cite{hayesArXiv2020} and the scanning tunneling microscopy measurements\cite{JiaoNature2020}.
We also point out that the spin-susceptibility-distributed signal in region~\mbox{I\hspace{-.1em}I} can be also explained by the properties of the superconductivity with the spin degree of freedom. 
It might be possible that the texture of $\bm{d}$-vector induces such a spatially modulated SC state in spin-triplet superconductors.

\section{Conclusion}
We performed $^{125}$Te-NMR measurements on a single-crystal UTe$_2$ to investigate the SC properties from a microscopic point of view.
A shoulder signal appeared inside the SC phase when $H \parallel b$.
The area fraction of the shoulder signal shows a maximum at $\sim 6$~T and vanishes above 10~T, although the superconductivity confirmed by the $\chi_{\rm AC}$ measurements survives to at least 14.5~T.
In contrast, an unexpected broadening without the shoulder signal was observed when $H~\parallel~c$ at a temperature below $T_{\rm c}$. This broadening is quickly suppressed by increasing $H$. 
In addition, the decrease in $K_b$ in the SC state starts to be small at around 7~T and almost zero at 12.5~T. 
This indicates that the SC spin state gradually changes with the application of $H$. 
Based on the experimental results, we constructed phase diagrams for both field directions.
The SC phase can be distinguished into four regions based on the character of the SC spin state.
In particular, an inhomogeneous low-$T$ SC state, where the high-$T$ and low-$T$ SC phases coexist, is expected to be realized in the low-$T$ and low-$H$ regions.
The low-$T$ SC phase originates from the presence of the spin degrees of freedom.
This finding places strong constraints on the SC pairing symmetry of UTe$_2$, and it also opens up new possibilities for the spin-triplet SC states.

\begin{acknowledgments}
The authors would like to thank M. Manago, J. Ishizuka, Y. Yanase, S. Fujimoto, Y. Maeno, and S. Yonezawa for their valuable discussions. 
This work was supported by the Kyoto University LTM Center, Grants-in-Aid for Scientific Research (Grant Nos. JP15H05745, JP17K14339, JP19K03726, JP16KK0106, JP19K14657, JP19H04696, JP20H00130, and JP20KK0061, and a Grant-in-Aid for JSPS Research Fellow (Grant No. JP20J11939) from JSPS.

\end{acknowledgments}

\end{document}